\documentclass[twocolumn,showpacs,showkeys,preprintnumbers,amsmath,amssymb,a4paper]{neel}

\usepackage{graphicx}
\usepackage{dcolumn}

\begin{document}

\DocReference{Submitted to Phys.Fluids}
\PageWeb{http://crtbt.grenoble.cnrs.fr/helio/}
\DateDerniereCompilation{Version : \today} 
\PSLogo{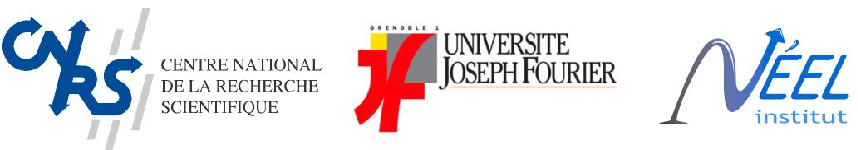}
\LogoHeight{1.7cm}


\title{Applicability of Boussinesq approximation in a turbulent fluid with constant properties}


\author{P.-E. Roche}
\affiliation{Institut NEEL, CNRS/UJF\\
BP 166, F-38042 Grenoble Cedex 9, France}


\date{\today}

\begin{abstract}
The equations of motion describing buoyant fluids are often simplified using a set of approximations proposed by J. Boussinesq one century ago. To resume, they consist in assuming constant fluid properties, incompressibility and conservation of calories during heat transport. Assuming fulfilment of the first requirement (constant fluid properties), we derive a set of 4 criteria for assessing the validity of the two other requirements in turbulent Rayleigh-B\'enard convection.
The first criterion $\alpha \Delta \ll 1 $ simply results from the incompressibility condition in the thermal boundary layer ($\alpha$ and $\Delta$ are the thermal expansion coefficient and the temperature difference driving the flow). The 3 other criteria are proportional or quadratic with the density stratification or, equivalently with the temperature difference resulting from the adiabatic gradient across the cell $\Delta_{h}$.
Numerical evaluations with air, water and cryogenic helium show that most laboratory experiments are free from such Boussinesq violation as long as the first criterion is fulfilled.
In ultra high Rayleigh numbers ($Ra>10^{16}$) experiments in He, one of the stratification criteria, scaling with $\alpha \Delta_{h}$, could be violated.  This criterion garanties that pressure fluctuations have a negligible influence both on the density variation and on the heat transfer equation through compression/expansion cycles.
Extrapolation to higher $Ra$ suggests that strong violation of Boussinesq approximation could occur in atmospheric convection.

\end{abstract}


\maketitle



\section{Introduction}

The buoyant motion of fluids combines the non-linear complexity of hydrodynamics with additional degrees of freedom brought in by the thermodynamics. The exact equations for the buoyant flows are therefore very difficult to tackle directly. In 1903, Joseph Boussinesq introduced a set of approximations and derived simplified equations for buoyant motion \cite{Boussinesq1903}. His three key approximations are~:\\
1) the fluid's properties are constant,\\
2) the flow is incompressible (except for a linear temperature dependence of density yielding the buoyant term),\\
3) the heat equation is not coupled to the flow mechanical energy (``conservation of calories")


Today, Boussinesq approximations are widely used to describe a large number of flows and to perform numerical simulations. A recent review presents an overview of works dedicated to the applicability of Boussinesq approximation and to its mathematical justification\cite{Zeytounian2003}. Oberbeck's name is occasionally associated with Boussinesq's to refer to the set of approximations presented above. As a side remark, it is worth noting that his convection model preserves compressibility and neglects all the time derivatives with respect to space derivatives \cite{Oberbeck1879}.
The resulting differential equations and physics significantly differs from the ones obtained by Boussinesq's.
\\

To be able to \textit{predict} if Boussinesq's approximations are applicable, specific quantitative criteria are necessary for each type of flow. The present work derives such criteria for Rayleigh-B\'enard (RB) turbulent convection\cite{Siggia1994,Kadanoff2001}, assuming constant properties of the fluid (first approximation satisfied). A vast literature is dedicated to this first approximation, in particular for high Prandtl numbers in connection with mantle convection (for example see \cite{Balachandar1993,Zhang1997,Manga1999,Sugiyama2007} and references within).  For the intermediate Prandtl numbers, recent progresses have been reported in the turbulent regime (for example see \cite{Wu1991,Frohlich1992,AhlersNOB006,AhlersNOB_PRL2007,BrownDraft} and references within).

Boussinesq second approximation (``incompressibility'') and third one (``conservation of calories'') consist in neglecting nearly half of the terms present in the exact equations of motion. When the magnitude of all the  terms can be estimated using typical scales of the flow, the applicability of the approximation can be predicted. In laminar RB convection, temperature and velocity gradients smoothly spread across the cell, which eases the choice of typical scales in the flow. A set of criteria for the applicability of Boussinesq approximation have been proposed in this low Rayleigh number ($Ra$) limit (for example see \cite{Gray1976,Tritton1988} and ref. within). In turbulent RB convection (high $Ra$), choosing the correct typical scales is less straightforward due to the flow complexity. In the present work, we attempt to do so using various experimental results and simple modelling.
We derive a new set of 4 criteria for the validity of Boussinesq approximation in a turbulent RB cell. In our derivation, we prioritise high Rayleigh numbers $Ra$ and intermediate Prandtl numbers ($Pr$)  fluids (say $Pr \sim 0.5-30$), as this range of parameters is relevant to many geophysical flows of air and water, industrial flows and high $Ra$ experiments with air, water, SF$_{6}$ and helium.
Using these criteria, we investigate the validity of Boussinesq approximations in experimental set-ups with water, air and helium. In particular, we show that one criteria is much more stringent than the others in most experiments, at least below $Ra \sim 10^{16}$ typically. At higher $Ra$, a second criteria could become relevant. We also  examine if deviations from Boussinesq approximation can explain the apparent incompatibility of heat transfer measurements among very high $Ra$ experiments. We find that it is not the case.

The next section recalls the exact equations of motion and singles out the different terms which have to be neglected to obtain Boussinesq's equations. In the third section, the magnitudes of these terms are compared with the retained terms by modelling turbulent convection. This yields a set of 4 criteria for the validity of Boussinesq approximation, which are summarised and discussed in the fourth section. The last section is a short conclusion.

\section{The Boussinesq equations}

The temperature, pressure and density  
 fields are decomposed into a constant (subscript ``0''), a stratification (subscript ``h'') and a residual ($\theta$, $p$ and $\delta \rho$) contribution~:
\begin{subequations}
\label{eq:definitions}
\begin{align}
\label{eq:temperature}
T& =T_{0}  +T_{h}+ \theta\\
\label{eq:pression}
P &= P_{0}+P_{h}+p\\
\label{eq:densite}
\rho& = \rho_{0}+  \rho_{h} +\delta \rho 
\end{align}
\end{subequations}

By definition, $T_{0}$ is the average of the top and bottom plates' temperatures. $T_{h}$ is the linearised \textit{adiabatic gradient} term defined by Eq.\eqref{eq:adiabatic}. The pressure $P_{0}+P_{h}$ and density $\rho_{0}+\rho_{h}$ fields satisfy the hydrostatic balance Eq.\eqref{eq:hydrostatic} and thermodynamic equilibrium Eq.\eqref{eq:thermostatic} of a static stratified fluid with a temperature distribution $T_{0} +T_{h}(z)$. At cell's mid-height ($z=0$), we set $\rho_{h}(0)=0$ and $P_{h}(0)=0$.
\begin{subequations}
\label{eq:static}
\begin{align}
\label{eq:adiabatic}
T_{h} &= - \frac{\alpha g T_{0}}{c_{p}} z \\
\label{eq:hydrostatic}
\frac{dP_{h}}{dz} &= - g \left( \rho_{0} + \rho_{h} \right) \\
\label{eq:thermostatic}
\frac{1}{ \rho_{0} + \rho_{h} } \frac{d\rho_{h}}{dz} &=  - \alpha \frac{dT_{h}}{dz} + \chi  \frac{dP_{h}}{dz}  
&= \frac{\alpha^2 g T_{0}}{c_{p}} - \chi g  \left( \rho_{0} + \rho_{h} \right)
\end{align}
\end{subequations}

\noindent where $\alpha$, $g$, $c_{p}$ and $\chi$ are the isobaric thermal expansion coefficient, the gravity acceleration, the isobaric heat capacity per unit mass and the isothermal compressibility coefficient.

The  residual terms $\theta$, $p$ and $\delta \rho$ account for fluctuations and for the inhomogeneities which are not captured by the stratification terms $T_{h}$, $P_{h}$ and $\rho_{h}$. For example, fluctuations are generated by turbulence and inhomogeneities result from the temperature forcing and confinement by the cell side walls. The potential temperature difference $\Delta$ and the adiabatic-gradient temperature difference $\Delta_{h}$ across a convection cell of height $h$ are defined as :
\begin{subequations}
\begin{align*}
\Delta &= \theta(-h/2)-\theta(h/2) \\
\Delta_{h} &=T_{h}(h/2)-T_{h}(-h/2)= \frac{\alpha g T_{0} h}{c_{p}}
\end{align*}
\end{subequations}
\noindent (the notation $\nabla^2$ will be used for the Laplacian operator to avoid confusion with the temperature difference $\Delta$). The mass, momentum and energy conservation equations are\,:
\begin{subequations}
\label{eq:conservation}
\begin{align}
\label{eq:cinematique}
\nabla \mathbf{v} & = -\frac{1}{\rho}\frac{D \rho}{Dt} \\
\label{eq:dynamique}
\rho \frac{D\mathbf{v}}{Dt} & = -\nabla P + \rho \mathbf{g} + \mathbf{F} \\
\label{eq:thermique}
\rho c_{p} \frac{DT}{Dt} &= k \nabla^2 T + \rho \Phi + \alpha T \frac{DP}{Dt}
\end{align}
\end{subequations}

\noindent where $D/Dt$ is the particle derivative and where $\mathbf{F}$ and $\Phi$ account for the viscous force and corresponding  dissipation and where $k$ is the thermal conductivity.

In this section, we assemble -step by step- a set of self consistent criteria which are sufficient to derive the Boussinesq equations from the conservation equations above. Since the successive approximations to derive Boussinesq equations are intricated, it is convenient to assume from the beginning that the density field is nearly homogeneous, or more precisely that the following two criteria are fulfilled (we omit the absolute values for better readability) :
\begin{subequations}
\label{eq:InequationsSmallDeviationsDensite}
\begin{align}
\label{eq:dhd0}
\rho_{h} &\ll \rho_{0}\\
\label{eq:ddd0}
\delta \rho &\ll \rho_{0}
\end{align}
\end{subequations}

\subsection{A preliminary thermodynamics relation}

Using the generalised Mayer relation and defining $\gamma=c_{p} / c_{v}$, we note that :
\begin{equation}
\label{relationThermo}
1 - \gamma^{-1} = \frac{\alpha ^2 T_{0}}{\rho c_{p} \chi}= \frac{\alpha \Delta _{h}}{\rho_{0} g \chi h}
\end{equation}

This equation relates the density stratification resulting from two different contributions~:  $\alpha \Delta _{h}$ is the thermal expansion associated with the adiabatic-gradient and $\rho_{0} g \chi h$ is the compressibility associated with the hydrostatic pressure. Since $\gamma$ is of order one or higher (critical point vicinity), this relation shows that the two contributions are comparable in strength.\\

%


%

\subsection{Momentum conservation}

Using the definitions Eq.\eqref{eq:pression} and \eqref{eq:densite}, the momentum conservation equation Eq.\eqref{eq:dynamique} can be written :
\begin{equation*}
\rho \frac{D\mathbf{v}}{Dt} = -\nabla p + \delta \rho \mathbf{g} + \mathbf{F} - \nabla P_{h} +  (\rho_{0}+\rho_{h}) \mathbf{g}
\end{equation*}

By definition of $P_{h}$, the last two terms cancel out.
Anticipating on the solenoidal condition $\nabla \mathbf{v}  =0$, the friction is restricted to the shear term $\eta \mathbf{\nabla}^2  \mathbf{v}$ which appears in Navier-Stokes equation (zero bulk viscosity).
Using the criteria Eq.\eqref{eq:InequationsSmallDeviationsDensite}, Boussinesq's momentum equation : 
\begin{equation}
\label{BoussiMomentum}
\rho_{0} \frac{D\mathbf{v}}{Dt}  =  -\nabla p - \rho_{0} \alpha \theta \mathbf{g} + \eta \mathbf{\nabla}^2  \mathbf{v}
\end{equation}

\noindent
is recovered if the extra criterion is fulfilled :
\begin{equation}
\label{eq:dilatation}
\delta \rho  \simeq - \rho_{0} \alpha \theta
\end{equation}

In the above equation, $\delta \rho $ is approximated at first order. We are assuming implicitly that higher order terms don't contain qualitative features which are significant for the flow dynamics. Such a "significant" feature is present -for example- in the viscous term (dissipation) and, consequently, it cannot be neglected with respect to the inertial term ${D\mathbf{v}}/{Dt}$, although the later is known to be larger in turbulent regimes.

\subsection{Mass conservation}

Using the definition Eq.\eqref{eq:densite}, the mass conservation Eq.\eqref{eq:cinematique} can be written :
\begin{equation*}
\nabla \mathbf{v}  = -\frac{1}{\rho}\frac{D \rho}{Dt} = -\frac{1}{\rho}\frac{D \delta \rho}{Dt}  -\frac{v_{z} }{\rho} \frac{d \rho_{h}}{dz}
\end{equation*}

Using the definition Eq.\eqref{eq:thermostatic}, the criteria Eq.\eqref{eq:InequationsSmallDeviationsDensite} and \eqref{eq:dilatation}, and the thermodynamics relation Eq.\eqref{relationThermo}, we get  : 
\begin{equation*}
\nabla \mathbf{v}  \simeq \alpha \frac{D \theta}{Dt}  + \frac{v_{z} \rho_{0} g \chi}{\gamma} 
\end{equation*}

Boussinesq's approximation requires that the velocity field is solenoidal ($\nabla \mathbf{v}  =0$), which is satisfied if each term in the right-hand side (RHS) of the previous equation can be neglected with respect to the typical velocity derivatives $\partial v_{i} / \partial x_{i}$ of the left-hand side (LHS) $\nabla \mathbf{v}$, where $i \in \{x, y, z\}$~:
\begin{subequations}
\label{eq:InequationsCinetique}
\begin{align}
\label{eq:condition.cinetique.a}
\frac{v_{z} \rho_{0}  g  \chi}{\gamma}  &\ll  \frac{ \partial v_{i} }{ \partial x_{i} }\\
\label{eq:condition.cinetique.b}
\alpha \frac{D \theta}{Dt} &\ll  \frac{ \partial v_{i} }{ \partial x_{i} }
\end{align}
\end{subequations}

\subsection{Energy conservation}

Using the definitions Eq.\eqref{eq:definitions} and \eqref{eq:static}, the energy conservation equation Eq.\eqref{eq:thermique} can be written :
\begin{equation}
\label{eq:EnergyConservation}
\rho c_{p} \frac{D\theta}{Dt} - k \nabla^2 \theta = \rho \Phi + \alpha T \frac{Dp}{Dt} + v_z g \alpha \left[ T_{0}  \rho -  T \left( \rho_{0} + \rho_{h} \right)    \right]
\end{equation}

Defining $\kappa=k / \rho c_{p}$ as the thermal diffusivity, Boussinesq's thermal equation (a ``conservation of calories'' equation) :
\begin{equation}
\label{eq.conservation_calories}
 \frac{D\theta}{Dt} 
 - \kappa \nabla^2 \theta 
 =0
 \end{equation}

\noindent is recovered if each term of the RHS of the equation Eq.\eqref{eq:EnergyConservation} can be neglected. A sufficient condition for this is found using the criteria Eq.\eqref{eq:InequationsSmallDeviationsDensite} and \eqref{eq:dilatation} and writing that $T_{0}$ provides a reasonable order of magnitude of $T$. We find the new criteria :
\begin{subequations}
\label{eq:InequationsThermique}
\begin{align}
\label{eq:condition.thermique.1}
{\Phi}/{c_{p}}  &\ll  \Theta^{\prime} \\
\label{eq:condition.thermique.2}
\frac{\alpha T_{0}}{\rho_{0} c_{p}} \frac{Dp}{Dt}  &\ll \Theta^{\prime} \\
\label{eq:condition.thermique.3}
\frac{v_z g \alpha}{ c_{p}} \left[ T_{h} +  \theta \left( 1 + \alpha T_{0} \right) \right]   &\ll \Theta^{\prime} 
\end{align}
\end{subequations}

where $ \Theta^{\prime} \in \{  \frac{\partial \theta}{\partial t} ,  v_{i} \frac{\partial \theta}{\partial x_{i}} , \kappa \frac{\partial^2 \theta}{\partial x_{i} ^2} \}$

\section{Criteria of applicability of the Boussinesq's approximation}

In this section, we evaluate the criteria  \eqref{eq:InequationsSmallDeviationsDensite},  \eqref{eq:dilatation}, \eqref{eq:InequationsCinetique}  and \eqref{eq:InequationsThermique} for the special case of high $Ra$ (turbulent convection) and intermediate $Pr$.

To check the first criterion Eq.\eqref{eq:dhd0}, the equations \eqref{eq:thermostatic} and \eqref{eq:hydrostatic} could easily be solved exactly but an order-of-magnitude solution for $\rho_{h}$ is enough for our purpose. Using also Eq.\eqref{relationThermo}, we easily find that the criterion Eq.\eqref{eq:dhd0} becomes\,:
\begin{equation}
\label{eq:a2}
\frac{ \rho_{0}  g h  \chi }{\gamma}  \ll 1\,\,\,\,\,\,\,\,\,\, \text{or} \,\,\,\,\,\,\,\,\,\,  \frac{\alpha \Delta_{h}}{\gamma -1} \ll 1
\end{equation}

To estimate the other criteria, we discriminate between two regions of the cell where the flow phenomenologies are known to differ : the bulk and the boundary layers near the top and bottom plates. Each criterion should be fulfilled in the two regions. Before proceeding with these two regions, it is useful to present order-of-magnitude fits of various quantities in the cell.

\subsection{Few orders of magnitude}

The  scalings of various flow quantities have been reported in the literature of turbulent convection, in particular in the Rayleigh-B\'enard cells. In this subsection, our aim is to present a few \textit{simple fits} that will be used to estimate and compare the \textit{orders of magnitude} of various contributions. We are not suggesting here that the chosen power law fits have an underlying physical meaning. We neglect the inhomogeneities within the boundary layers and within the bulk of the flow, which result from finite size effects e.g. confinement of the large scale circulation. 

Below $Ra=10^{12}$ and for $0.7<Pr<30$, all reported $Nu(Ra)$ data are in reasonable agreement (for studies of both $Ra$ and $Pr$ in this range, see for example  \cite{Verzicco1999,Kerr2000,Ahlers2001,RocheEPL2002,Hartlep2005}). We will use the fit :
\begin{equation}
\label{eq:Nu}
(Ra<10^{12})\,\,\,\,\,\,Nu \simeq 0.065 Ra^{1/3}
\end{equation}
The $1/3$ exponent chosen in this fit is slightly larger than the exponents fitted in most studies (which are closer to 0.30-0.32), even when the temperature drop in plates is compensated \cite{Verzicco_plate}. But this simple fit still provides good numerical estimates for our purpose.

Above $Ra=10^{12}$, the reported $Nu(Ra)$ dependence fall into two sets : data from 2 experiments roughly follows the scaling of Eq.\eqref{eq:Nu} (\cite{Wu1992,Niemela2000}) while the others exhibit a steep increase of the $Ra$ exponent from 0.31 up to typically 0.38  \cite{Chavanne1997,Niemela2003,RochePoF2005,Niemela:2006}.
 The enhanced-heat-transfer law can be fitted by\,:
\begin{equation*}
(10^{12}<Ra<10^{14} ; Pr \simeq 10)\,\,\,\,\,\,Nu \simeq 0.015 Ra^{0.38}
\end{equation*}
The apparent incompatibility between the two sets of data is not understood and motivates present research (see also the numerical results \cite{Kenjeres2002,Amati2005}). Although all the data showing a heat transfer enhancement (including \cite{RochePRE2001}) are consistent with the ``ultimate regime'' predicted by R. Kraichnan \cite{Kraichnan1962}, an extensive characterisation of this new regime is still missing. Therefore, we will not attempt to derive specific Boussinesq criteria for it in this paper. In the Discussion section, we will come back on this point to discuss very high $Ra$ experiments.\\

A large scale circulation or ``wind'' is known to occur in Rayleigh-B\'enard cells. The strength of the  l.s.c velocity $v_{lsc}$  can be characterised by the fit :
\begin{equation}
\label{eq:Relsc}
Re = \frac{h v_{lsc} }{ \nu}  \simeq 0.15 Ra^{1/2} Pr^{-3/4}
\end{equation}

\noindent
which is 
a compromise between studies from various groups \cite{Ashkenazi,Chavanne2001,Lam2002,QiuPRE2002,QiuPOF2004,Niemela:2006,Brown_Re2007} conducted with various operational definitions of $Re$ for $0.7<Pr<30$, based on local velocity or turn over time measurements \cite{Sun_PRE2005}.
An alternative estimate of the typical velocity in the bulk of the flow can be obtained from the rms velocity fluctuations. Comparing the fit of $Re$ (above) with rms velocity data from the literature \cite{Daya2002,Verzicco2003,Lam2002,QiuPOF2004} 
we note that the scaling of the root-mean-squared (rms) velocity is approximately the same as the scaling of the large scale circulation velocity, and that the magnitudes are comparable (rms velocity is typically $40\%$ of $v_{lsc}$). We will therefore consider only a single characteristic velocity at large scales.  We should mention that, in the literature, the $Ra$ dependence of $Re$ has been fitted with effective exponents as low as $0.40$ \cite{Lam2002} (and even $0.36 \pm 0.05$ at some location in a square cell \cite{Daya2001} ) and as high as 0.65 (in the horizontal plane \cite{QiuPOF2004}).\\

Many consistent studies report scalings of  the rms temperature fluctuations in the bulk of the cell versus $Ra$ (for example see \cite{Wu1992,Xia1997,Niemela2000, Kerr2001,Du2001}) but, to our knowledge, only Daya and Ecke \cite{Daya2002} reports both a $Ra$ and $Pr$ dependence. The fit below is derived from Daya and Ecke original data. We should mention that the reported $Pr$ dependence is poorly fitted by a power law but that such a fit still provides a fair order of magnitude for our purpose. $\theta_{rms} ^{\star}$ is defined as the rms temperature fluctuations normalised by the temperature difference $\Delta$ across the cell.
\begin{equation}
\label{Daya}
 \theta_{rms} ^{\star}   \simeq    0.3 Ra^{-0.13} Pr ^{-0.5}   
\end{equation}

Finally, we will need an estimate of the time derivative $\partial \theta / \partial t$  in the bulk of the flow. The references
\cite{WuTHESE, Procaccia1991} report corresponding measurements over a large range of $Ra$ ($5.10^8<Ra<10^{15}$) for  Prandtl numbers of order 1-10 (see also \cite{Du2001}). They interpret a transition occurring near $Ra \sim 10^{11}$ with a change in power laws. Nevertheless, within a factor 2, we can still re-fit all their data with the following power law, which is enough for our purpose  (the Prandtl number dependence is unknown) :
\begin{equation*}
\partial \theta / \partial t \simeq 0.7 Ra^{0.4}      \frac{ \Delta \kappa}{h^2}
\end{equation*}

In the bulk of the flow, we expect advection to be the most effective heat transport mechanism, which implies $\partial \theta / \partial t \simeq \mathbf{v} \nabla \theta$. The relation shows that we can also estimate the time derivative from $v \nabla \theta$, taking $v \simeq \nu Re /h$ and using recent temperature-gradient measurements for $10^9 < Ra < 10^{10}$ and $Pr \simeq 5.4$ \cite{He2007}.  Averaging the measured gradient from the various locations inside the cell\cite{He2007}, we obtain the fit :
\begin{equation*}
\partial \theta / \partial t \simeq \mathbf{v} \nabla \theta  \sim v | \nabla \theta |   \simeq  70 Ra^{-0.165} Re     \frac{ \Delta \nu}{h^2}  
\end{equation*}

\noindent
It is satisfactory that both estimations are in reasonable agreement (magnitude and $Ra$ exponent). We will favour the first fit because its has been validated over a wider range of $Ra$. It is convenient to define a dimensionless derivative $\theta _{\partial t} ^{\star}$ as :
\begin{equation}
\label{DthetaDtAdim}
\theta _{\partial t} ^{\star} = \frac{\partial \theta}{ \partial t}   \frac{h^2}{ \Delta \kappa}  \simeq 0.7 Ra^{0.4} x    
\end{equation}

\noindent
where $x=(Pr/5)^X$ accounts for the unknown Prandtl dependence through the unknown parameter $X$. For intermediate $Pr$, we will assume that $x$ is of order 1.

\subsection{Boundary layer region}

The present knowledge of scaling in the boundary layers is not as extended as in the bulk, in particular due to the complex interplay between the thermal and the velocity boundary layers (for recent progresses, see  \cite{DuPuitsJFM2007,VanReeuwijkTHESE,Maystrenko2007}). Nevertheless, it is well established that each boundary layer concentrate nearly half of the total temperature drop  $\Delta$ over a typical distance of order $\lambda = h/2Nu$. An internal velocity scale $v_{bl}$ can be obtained writing that total heat flux experiences a cross-over within the thermal boundary layer from a molecular diffusion transport (right on the plate) to a convective transport at a distance $\lambda$ from the plate, which gives $v_{bl} = \kappa / \lambda = 2Nu. \kappa / h$. An external velocity scale is also imposed on the thermal boundary layers by the large scale circulation present in the bulk of the cell $v_{lsc}  = Re  . \nu / h$. Using the estimates of $Nu$ and $Re$ given previously, we note that $v_{lsc} > v_{bl}$. In the following, we will choose the velocity scale which gives the more stringent criteria (``sufficient conditions'').
Using these temperature, length and velocity scales, we evaluate below the criterion \eqref{eq:ddd0},  \eqref{eq:dilatation}, \eqref{eq:InequationsCinetique}  and \eqref{eq:InequationsThermique} .

Local thermodynamic equilibrium implies the differential equation :
\begin{equation*}
d\rho =  - \rho \alpha dT+ \rho \chi dP
\end{equation*}

Subtracting the stratification contribution \eqref{eq:thermostatic} from this local equilibrium, the criteria Eq.\eqref{eq:ddd0} and \eqref{eq:dilatation} are found to be equivalent to  :
\begin{subequations}
\label{eq:critereDensiteEquivalent}
\begin{align}
\alpha \theta &\ll 1\\
\chi p &\ll \alpha \theta
 \end{align}
\end{subequations}

In the boundary layer, we have $\theta \sim \Delta /2$. The order of magnitude of $p$ is assessed taking the divergence of Eq.\eqref{BoussiMomentum} and using the solenoidal property of the velocity field. We find a Poisson equation for the pressure :
\begin{subequations}
\label{Poisson}
\begin{align}
{\nabla ^2 p} &= - \rho_{0} \mathbf{ \nabla } \left[   \left( \mathbf{v} \nabla \right) \mathbf{v} \right] + \alpha \rho_{0} g \frac{\partial \theta}{\partial z}  
 \end{align}
\end{subequations}

\noindent
The first term of the RHS gives the usual dynamical pressure of order :
\begin{equation*}
p \sim  \frac{\rho_{0}}{2} (  {v_{bl}}^2 + {v_{lsc}}^2 ) \sim  \frac{\rho_{0}}{2}  {v_{lsc}}^2 
\end{equation*}

\noindent
and the second RHS term, a buoyant source of pressure variation, can be estimated by integration as :
\begin{equation*}
p \sim  \frac{g \rho_{0} \alpha \Delta \lambda 
 }{4}
\end{equation*}

Substituting the expressions $v_{lsc} = \nu Re /h$ and $\lambda = h / 2Nu$ and using the fits for $Nu$ and $Re$, we find that the buoyant contribution to the pressure is smaller than the kinematic contribution. Neglecting the former and using Eq.\eqref{relationThermo}, both criteria Eq.\eqref{eq:critereDensiteEquivalent} becomes, up to a factor 2 :
\begin{subequations}
\label{eq:critereDensiteBL}
\begin{align}
\label{eq:b}
\alpha \Delta &\ll 1\\
\label{eq:stratTemp}
 \frac{Pr Re^2}{Ra} \frac{\alpha \Delta_{h}}{1-\gamma^{-1}}    &\ll 1
 \end{align}
\end{subequations}

The second criterion Eq.\eqref{eq:stratTemp} is comparable or less stringent than Eq.\eqref{eq:a2}, so we can drop it.\\

In the boundary layer, the next criterion Eq.\eqref{eq:condition.cinetique.a} is estimated as :
\begin{equation*}
\frac{v \rho_{0} g \chi}{\gamma} \ll \frac{v}{\lambda}
\end{equation*}
 
 After simplification by the typical velocity $v$ and using the expression of  $\lambda$, we find that the above condition is  less stringent than the criterion Eq.\eqref{eq:a2}. We will therefore ignore it too.\\

To evaluate the criterion Eq.\eqref{eq:condition.cinetique.b} in the boundary layer,  the particle derivative
of the temperature is estimated with a diffusion process (heat is indeed transferred between the plates and the fluid by diffusion in the boundary layers) :
\begin{equation}
\label{eq:detached}
\frac{D\theta}{Dt} \simeq  \frac{ \Delta /2 }{ \lambda^2 / \kappa}
\end{equation}

\noindent
and the space derivative of the velocity is estimated from the typical internal velocity as $v_{bl} / \lambda$. The external velocity $v_{lsc}$ is not chosen because it would give a more loose condition. Substituting the expressions of ${D\theta}/{Dt}$, $v_{bl}$ and $\lambda$ in the criterion Eq.\eqref{eq:condition.cinetique.b} gives criterion Eq.\eqref{eq:b} again.\\

We now consider the three criteria Eq.\eqref{eq:InequationsThermique}  associated with the energy equation. The three RHS term $\Theta^{\prime}$ of these inequalities can be estimated with the characteristic scales used above. All turn out to have the same typical magnitude, as expected from the chosen boundary layer model :
\begin{equation*}
\Theta^{\prime}  \simeq \frac{\partial \theta}{\partial t} \simeq  v_{i} \frac{\partial \theta}{\partial x_{i}} \simeq \kappa \frac{\partial^2 \theta}{\partial x_{i} ^2} \simeq \frac{\kappa \Delta}{2 \lambda^2}
\end{equation*}

We estimate the LHS of the inequalities Eq.\eqref{eq:InequationsThermique} using $v_{lsc}$ as the typical velocity (``sufficient condition'').
\begin{align*}
\frac{\Phi}{c_{p}}  &\lesssim \frac{\nu}{c_{p}} \left( {\frac{v_{lsc}}{\lambda}} \right) ^2 \\
\frac{\alpha T_{0}}{\rho_{0} c_{p}} \frac{Dp}{Dt}   &< \frac{\alpha T_{0}}{\rho_{0} c_{p}} \frac{\rho_{0} {v_{lsc}}^2/2 }{\lambda/ v_{lsc}} \\
\frac{v_z g \alpha}{ c_{p}} \left[ T_{h} +  \theta \left( 1 + \alpha T_{0} \right) \right]   &\lesssim  \frac{v_{lsc} g \alpha}{ c_{p}} \left[ \frac{\Delta_{h}}{2} +  \frac{\Delta}{2} \left( 1 + \alpha T_{0} \right) \right] 
 \end{align*}

Substituting the expressions of $v_{lsc}(Re)$ and $\lambda (Nu)$ in the previous equations, we find up to prefactors of order unity :
\begin{subequations}
\label{eq:InequationsThermique.bl}
\begin{align}
\label{eq:InequationsThermique.bl1avantFusion}
\frac{\Delta_{h}}{T_{0}}   \times  \frac{2 Pr^2 Re^2}{Ra}  &\ll 1 \\
\label{eq:InequationsThermique.bl2NewEject}
\alpha \Delta_{h}      \times   \frac{Pr^2 Re^3}{2 Ra Nu}  &\ll 1 \\
\label{eq:InequationsThermique.bl3New}
\frac{\Delta_{h}}{T_{0}} \left( \frac{\Delta_{h}}{\Delta} + \alpha T_{0} + 1 \right)     \times    \frac{Pr Re }{4 Nu^2}     &\ll 1
 \end{align}
\end{subequations}

The second condition \eqref{eq:InequationsThermique.bl2NewEject} is comparable or more stringent than Eq.\eqref{eq:a2} and will replace it. The  $\alpha T_{0}$ term in the parenthesis of Eq.\eqref{eq:InequationsThermique.bl3New} gives a condition comparable or less severe than \eqref{eq:InequationsThermique.bl2NewEject} and can therefore be dropped. The  unity term in the parenthesis gives a condition that can be compared to Eq.\eqref{eq:InequationsThermique.bl1avantFusion} and it turns out to be typically up to 10 times more stringent for low $Ra$ ($Ra \sim 10^7$), comparable for $Ra \sim 10^{13}$, and less stringent above. We will come back on these 2 conditions later.
The ${\Delta_{h}}/{\Delta}$ term of Eq.\eqref{eq:InequationsThermique.bl3New} gives the condition :
\begin{equation}
\label{eq:InequationsThermique.bl3nettoyeeNew}
\frac{\Delta_{h}^2}{T_{0} \Delta}  \times    \frac{Pr Re }{4 Nu^2}     \ll 1
\end{equation}


\subsection{Bulk region}

In the bulk, for the same reasons as in the boundary layers, the criteria Eq.\eqref{eq:ddd0} and \eqref{eq:dilatation} are equivalent to  :
\begin{align*}
\alpha \theta &\ll 1\\
\chi p &\ll \alpha \theta
 \end{align*}

The first condition is weaker than the criteria Eq.\eqref{eq:b} since the magnitude of temperature fluctuations in the bulk  is significantly smaller than the total temperature drop across the cell   ($\theta_{rms}^{\star} \ll 1)$.

To assess the second condition, the magnitude of the pressure $p$ is estimated using the Poisson equation Eq.\eqref{Poisson}. The buoyant contribution is estimated by analogy with electrostatic, $\alpha \rho_{0} g  {\partial \theta}/{\partial z} $ being the  pressure source with a negligible remote contribution from the boundary layers\footnote{In very compressible fluids (therefore non-Boussinesq), remote contribution from boundary layers could become significant due to the so called ``piston effect''} and a local contribution from the bulk.
\begin{equation*}
p \lesssim  \frac{\rho_{0}}{2} v^2 +   \alpha  \rho_{0} g    |  \frac{\partial \theta}{\partial z}|  l^2
\end{equation*}

\noindent
where $l$ is a correlation length of the temperature gradient ${\partial \theta}/{\partial z}$ in the bulk.
The buoyant contribution $ |  \frac{\partial \theta}{\partial z}|  l^2$ can be estimated from the temperature rms fluctuation $\Delta \theta_{rms}^{\star}$,  the typical temperature derivative $ {\partial \theta}/{\partial z_{i}}$, and using the inequality 
\begin{equation*}
|  \frac{\partial \theta}{\partial z}|  l^2 < |  \frac{\partial \theta}{\partial z}|  h l \sim |  \frac{\partial \theta}{\partial z}|  h \frac{ \Delta \theta_{rms}^{\star} }{ { |\partial \theta}/{\partial z}  |} \sim h \Delta \theta_{rms}^{\star} 
\end{equation*}


Estimating the typical velocity as $v  = \nu Re / h$ and using the fit of $Re$ and $\theta_{rms}^{\star}$,  we find that the buoyant contribution to the pressure variation is comparable or smaller than the kinematic term for $Ra>10^8$. We will ignore it. A sufficient condition for the criteria $\chi p \ll \alpha \theta$ in the bulk is therefore :
\begin{equation}
\label{eq:stratNew}
\frac{\alpha \Delta_{h}}{1-\gamma^{-1}}   \times  \frac{Pr Re^2 }{2 Ra \theta_{rms}^{\star}}     \ll 1
\end{equation}

This new criterion is comparable to Eq.\eqref{eq:InequationsThermique.bl2NewEject}. We choose the new one because it is derived from tighter estimates.

To estimate the criteria Eq.\eqref{eq:condition.cinetique.a}, we take $ v_{z} \sim \nu Re  / h$ as the typical vertical velocity. The typical derivative $\partial v_{i} / \partial x_{i}$ is estimated assuming that the velocity fluctuations in the bulk of the flow are driven by the large scale circulation on a large scale $h$ and follows a Kolmogorov-like cascade. Then,  the energy injected into cascade is related to the viscous dissipation $\Phi$, which is proportional to the square of velocity spatial derivatives (\cite{MoninYaglom,Grossmann2000}) :
\begin{equation}
\label{eq:Kolmo}
\frac{v_{lsc}^3}{h} = Re^3  \times   \frac{ \nu^3}{h^4} \simeq \Phi \sim \nu \left( \frac{\partial v_{i}}{\partial x_{j}} \right)^2 
\end{equation}

from which we get the estimate :
\begin{equation*}
\frac{\partial v_{i}}{\partial x_{i}}  \sim  Re^{3/2}     \frac{ \nu}{h^2}
\end{equation*}

Using those estimates and Eq.\eqref{relationThermo}, it is straightforward to show that  Eq.\eqref{eq:condition.cinetique.a} in the bulk is less stringent than Eq.\eqref{eq:a2}. We can therefore ignore this condition. As a comment, the LHS of Eq.\eqref{eq:condition.cinetique.a} has been estimated using the magnitude of the velocity at large scales while the RHS has been estimated at small scales where the derivative is known to be larger. The solenoidal condition Eq.\eqref{eq:condition.cinetique.a} is nevertheless still accounted for at large scale only, as can be checked estimating the RHS as $v_{lsc}/h$  (one gets Eq.\eqref{eq:a2}).\\

To estimate the next criterion Eq.\eqref{eq:condition.cinetique.b}, we need an upper bound for the particle derivative $D\theta / Dt$. We can consider the relaxation by molecular diffusion of a $\lambda$-thick plume of excess temperature $\Delta /2$. The result is the same as in the boundary layer : Eq.\eqref{eq:detached}. The criterion Eq.\eqref{eq:condition.cinetique.b} becomes in the bulk :
\begin{equation*}
   \frac{2 Nu ^{2}}{Pr Re^{3/2}}  \alpha \Delta  \ll 1
\end{equation*}

Using typical fits $Re(Ra)$, $Nu(Ra)$, we find that the previous inequality is less severe than \eqref{eq:b} and can be ignored.\\

We now consider the three criteria Eq.\eqref{eq:InequationsThermique}  associated with the energy equation. Boussinesq thermal equation Eq.\eqref{eq.conservation_calories} states that the advection of temperature inhomogeneties is only affected by the molecular thermal diffusion. In the boundary layers, we showed that these two transport processes have the same efficiency. In the bulk, we expect advection to become the dominant process, except in some localised regions (on the thermals interface) where diffusion could be as efficient as in the boundary layers. To estimate the three inequalities Eq.\eqref{eq:InequationsThermique}, we will require that each LHS term is significantly smaller than the advection term $\mathbf{v}  \nabla \theta \simeq \partial \theta / \partial t  = \theta_{\partial t }^{\star} \Delta \kappa / h^2 $ of the RHS. It should be noted that the molecular diffusion term in the thermal equation provides a cut-off (or ``dissipation'') mechanism for temperature fluctuations at small scales and that we are not aiming to compare the magnitude of this cut-off term with the LHS terms of  Eq.\eqref{eq:InequationsThermique}.


To estimate the LHS of the first thermal criterion Eq.\eqref{eq:condition.thermique.1}, Eq.\eqref{eq:Kolmo} provides an order of magnitude of the viscous dissipation $\Phi$
which is found consistent with the Direct Numerical Simulation (DNS) results of Verzicco \cite{Verzicco2003}. The criterion Eq.\eqref{eq:condition.thermique.1} in the bulk becomes Eq.\eqref{eq:condition.thermique.1.BulkavantFusion}.

To estimate the next criterion Eq.\eqref{eq:condition.thermique.2}, we assume that pressure and temperature fluctuates with similar typical time scales. The derivatives $Dp / Dt$ and $\partial \theta / \partial t$ of Eq.\eqref{eq:condition.thermique.2} are replaced with their integrated value $p\sim \rho_{0} v^2/2$ and $\theta \sim \Delta  \theta_{rms}^{\star}$. We find Eq.\eqref{eq:condition.thermique.2.BulkNew}.

To estimate the LHS of the third thermal criterion Eq.\eqref{eq:condition.thermique.3}, we take $v_{z} \sim Re.\nu  /h$, $\theta \sim \Delta  \theta_{rms}^{\star}$ and $T_{h} \sim \Delta_{h}/2$ to find Eq.\eqref{eq:condition.thermique.3.BulkNew}.
\begin{subequations}
\label{eq:InequationsThermiqueBulkNew}
\begin{align}
\label{eq:condition.thermique.1.BulkavantFusion}
 \frac{\Delta _{h}}{T_{0} }   \times   \frac{ Pr^2 Re^3} {Ra  \theta _{\partial t} ^{\star} }  &\ll 1 \\
\label{eq:condition.thermique.2.BulkNew}
\alpha \Delta_{h}    \times   \frac{Pr Re^2 }{2 Ra \theta_{rms}^{\star}}  & \ll 1 \\
\label{eq:condition.thermique.3.BulkNew}
 \frac{ \Delta _{h} }{  T_{0}  } \left[ \frac{\Delta _{h}}{2 \Delta \theta_{rms}^{\star} } +  \alpha T_{0} +  1 \right]  \times   \frac{ Pr Re \theta_{rms}^{\star}} { \theta _{\partial t} ^{\star} }   & \ll 1
\end{align}
\end{subequations}

Using the typical orders of magnitude of $Re$, $\Delta \theta_{rms}^{\star}$ and $\theta _{\partial t} ^{\star}$, we now compare these three thermal conditions Eq.\eqref{eq:InequationsThermiqueBulkNew} with previous ones. The first condition  is less stringent than Eq.\eqref{eq:InequationsThermique.bl3New} (unity term) at low $Ra$ and
more stringent than Eq.\eqref{eq:InequationsThermique.bl1avantFusion} at high $Ra$. We choose to drop the condition Eq.\eqref{eq:InequationsThermique.bl1avantFusion} and to merge Eq.\eqref{eq:InequationsThermique.bl3New} (unity term only) and Eq.\eqref{eq:condition.thermique.1.BulkavantFusion} into :
\begin{equation}
\label{eq:condition.thermique.1.BulkapresFusion}
\frac{\Delta_{h}}{T_{0}}   \times  \left(  \frac{Pr Re }{4 Nu^2}  +   \frac{ Pr^2 Re^3} {Ra  \theta _{\partial t} ^{\star} }     \right)    \ll 1
\end{equation}
Up to a numerical factor $(1-\gamma^{-1})$ of order one, the second thermal condition Eq.\eqref{eq:condition.thermique.2.BulkNew} is equivalent to the criterion \eqref{eq:stratNew}.
We now consider one by one each term in the brackets of Eq.\eqref{eq:condition.thermique.3.BulkNew}. The unity term gives a condition which is less stringent than Eq.\eqref{eq:condition.thermique.1.BulkapresFusion}. The $\alpha T_{0}$ term  gives a condition less stringent than Eq.\eqref{eq:condition.thermique.2.BulkNew} and can be ignored. The ${\Delta _{h}}/{\Delta \theta_{rms}^{\star} } $ term  gives a condition comparable, or more stringent for high $Ra$, than Eq.\eqref{eq:InequationsThermique.bl3nettoyeeNew}. We therefore keep the former one and re-write it alone as
\begin{equation}
\label{eq:condition.thermique.3.BulkNettoyee}
\frac{\Delta_{h}^2}{T_{0} \Delta}  \times    \frac{Pr Re }{2 \theta _{\partial t} ^{\star}}     \ll 1
\end{equation}

\subsection{Summary of the criteria and discussion}

The set of the most stringent criteria for the applicability of Boussinesq approximation leaves us with only four sufficient conditions  Eq.\eqref{eq:b}, \eqref{eq:stratNew}, \eqref{eq:condition.thermique.1.BulkapresFusion} and \eqref{eq:condition.thermique.3.BulkNettoyee}, which are recalled below, up to a factor 2.
\begin{subequations}
\label{eq:summary}
\begin{align}
\label{eq:summaryB}
\alpha \Delta  &\ll 1 \\
 \label{eq:summaryJnew}
 \alpha \Delta_{h}    \times  \frac{Pr Re^2}{Ra  \theta^{\star}_{rms}}  & \ll 1 \\ 
 \label{eq:summaryInew}
 \frac{\Delta _{h}}{T_{0} }   \times    \left(   \frac{ Pr^2 Re^3} {Ra  \theta _{\partial t} ^{\star} }   +  \frac{Pr Re }{4 Nu^2}   \right)  &\ll 1 \\
\label{eq:summaryL}
\frac{\Delta_{h}^2}{T_{0} \Delta}  \times    \frac{Pr Re }{2 \theta _{\partial t} ^{\star}}    & \ll 1
\end{align}
\end{subequations}
\noindent
and we recall that :
\begin{equation*}
\alpha \Delta _{h} ={\alpha^2 g T_{0} h} / {c_{p}} = \rho_{0} g \chi h (1 - \gamma^{-1} ) \sim  \rho_{0} g \chi h
\end{equation*}

These 4 criteria are a first outcome of this paper. Although the choices of these specific criteria is not free from some arbitrariness, the order of magnitude of the LHS of these inequalities should be fairly robust to these choices.\\

Physically, the first criterion is required for incompressibility in the boundary layer. The other criteria are more difficult to relate to a specific physical effect. For example, the complete heat equation has a term accounting for the thermo-mechanical energy coupling which directly results from stratification \footnote{This thermo-mechanical effect is already significantly compensated  in the Rayleigh-B\'enard geometry by the use of potential temperature. This compensation can be more difficult to implement in other set-ups such as differentially heated cavity with adiabatic horizontal walls \cite{Pons2005a,Pons2005b}}. This term -which appears on the LHS of Eq.\eqref{eq:condition.thermique.3}- gives contributions proportional to $\alpha \Delta_{h}$, $\Delta_{h} /T_{0}$ and ${\Delta_{h}^2}/{T_{0} \Delta}$. The fourth criterion above accounts for the ${\Delta_{h}^2}/{T_{0} \Delta}$ contribution alone. The third criterion above includes the $\Delta_{h} /T_{0}$ contribution but also another condition necessary to neglect the viscous dissipation with respect to advected heat flux. The second criterion above 
allows to neglect variations of density due to the pressure variations, versus those due to temperature variations (Eq.\eqref{eq:stratNew}. It also allows to neglect the cooling/heating associated with the pressure variations experienced by a fluid particle in the heat transport equation (Eq.\eqref{eq:InequationsThermique.bl2NewEject} and Eq.\eqref{eq:condition.thermique.2.BulkNew}).\\

These 4 criteria don't allow to assess \textit{a-priori} the fulfilment of the Boussinesq approximation for given $Ra$ and $Pr$ since it requires some knowledge of $Re$, $\theta _{\partial t} ^{\star}$ and $\theta_{rms}^{\star}$ . A further step in the analysis consists in replacing these quantities by fits versus $Ra$ and $Pr$. For example, using the fits presented previously, the 4 previous criteria becomes :
\begin{subequations}
\label{eq:summaryExplicit}
\begin{align}
\label{eq:summaryBexplic}
\alpha \Delta  &\ll 1 \\
 \label{eq:summaryJexplic}
0.1 Ra^{0.13}  \left( \alpha \Delta_{h}   \right)   & \ll 1 \\
 \label{eq:summaryIexplic}
  \left[ \frac{Ra^{0.1} Pr^{-0.25} x}{200}  + 10 Ra^{-0.17} Pr^{0.25} \right] \left(  \frac{\Delta _{h}}{T_{0} } \right)    &\ll 1 \\
\label{eq:summaryLexplic}
0.1 Ra^{0.1} Pr^{0.25} x   \left(  \frac{\Delta_{h}^2}{T_{0} \Delta} \right)    & \ll 1
\end{align}
\end{subequations}

\noindent where we rounded prefactors and exponents to recall that the above inequalities are derived from order-of-magnitude fits. We recall that the order of magnitude of $x$ is taken as unity for $Pr$ of order 1-10.

These equations highlight parameters having a direct influence on the validity of the Boussinesq approximation. For example, the dependence versus $Ra$ is very weak. The equations Eq.\eqref{eq:summaryExplicit} indicates that high $Ra$ are weakly unfavourable but the uncertainty in the fits' exponents prevents from drawing precise conclusions on  this weak dependence. For given $Ra$ and $Pr$, we found general features. Tall cells are unfavourable because of stratification (linear and quadratic dependence versus $h$ coming from $\Delta_{h}$). Close proximity to the critical point is unfavourable too since $\alpha$ has a strong divergence in a region where $\Delta_{h}$ stays approximately constant. Finally, large temperature differences $\Delta$ have opposite influence on the criterion Eq.\eqref{eq:summaryBexplic} (unfavourable) and on the criterion Eq.\eqref{eq:summaryLexplic} (favourable).  In practice all these parameters can't all be varied independently.\\



Numerical applications are now proposed for a few typical laboratory Rayleigh-B\'enard cells. Let's first consider a water cell of height $h=1\,m$, mean temperature $T_{0}=50^{\circ}C$ and driven with $\Delta = 30^{\circ}C$ (near atmospheric pressure). Such cells are being studied -for example- in physics labs in Lyon, Hong-Kong and Santa Barbara (for example see  \cite{Chilla2004_TiltLongRelax,Sun2005,Nikolaenko2005}). We found that the criteria Eq.\eqref{eq:summaryExplicit} are all fulfilled. The most stringent one is by far Eq.\eqref{eq:summaryBexplic}, which is nevertheless still valid with 2 decades separation between the two sides of the inequality. The three other criteria are fulfilled with at least 5 decades of separation.\\

In $h=8\,m$ air cell at atmospheric pressure, such as the one operated in Ilmenau (for example see  \cite{duPuits_PRE2007}), we also find that the first criterion  Eq.\eqref{eq:summaryBexplic} is the most stringent. For $\Delta = 50^{\circ}C$ and $T_{0}=50^{\circ}C$, we find a ration 6  separation between the two sides of this inequality. The other criteria are fulfilled with 2.5 or more decades of separation.\\

Although atmospheric convection differs from RB convection of dry air, it is interesting the examine the criteria Eq.\eqref{eq:summaryExplicit} for $h=10\,km$, $\Delta = 50\,C$ and $T_{0}=0^{\circ}C$. Using dry air properties, we find that Eq.\eqref{eq:summaryJexplic} is violated by more than one decade and than the LHS of Eq.\eqref{eq:summaryIexplic} is close to 1, indicating that the Boussinesq approximation is not valid in this academic case. Surely these numbers should be taken with much caution because they rely on the extrapolation up to $Ra\simeq 2.10^{21}$ of fits derived at much lower $Ra$ and probably in a different turbulent regime.\\

The case of Helium cells, such as those being operated in Grenoble and Trieste 
cannot be addressed with the same general procedure as above because, from one run to another,  the temperature and the fluid density are varied and the parameter $\alpha$ does change accordingly (while $\Delta _{h} / h$ remains close to $2.10^{-3}$ - $3.10^{-3}\,K m^{-1}$). 
 Besides, the modelling presented in this paper ignores the possibility of occurrence of the Kraichnan's ultimate regime of convection \cite{Kraichnan1962}, while the interpretation of some high $Ra$ data is done according to this regime.
These words of caution written, we can still get guidelines from the derived criteria. We numerically find that the two first criteria Eq.\eqref{eq:summaryBexplic} and Eq.\eqref{eq:summaryJexplic} are the most stringent ones, while the two others are fulfilled by at least 3 orders of magnitude for cells of height $20\,cm$ and $1\,m$
Therefore, we focus below on the validity of the second criterion for experimental data fulfilling the first one.

\begin{figure}
\includegraphics[width=8.5cm]{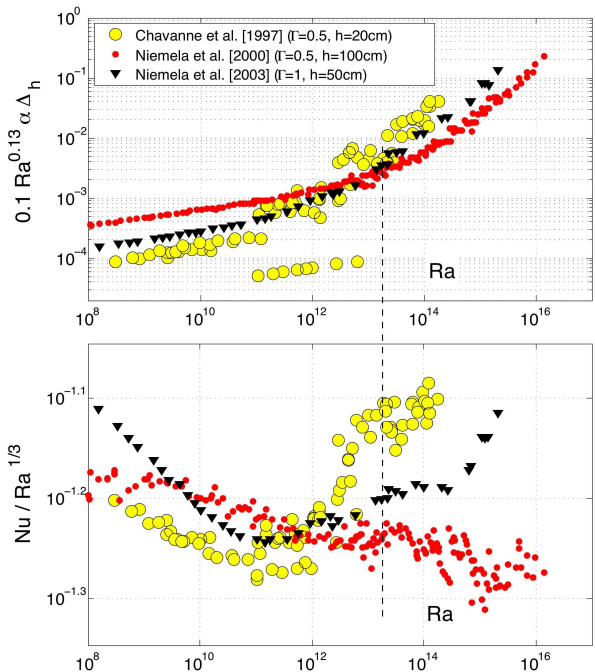}
\caption{Data from three cryogenic He experiments\cite{Chavanne1997, Niemela2000,Niemela2003}, restricted to those fulfilling $\alpha \Delta < 0.2$.  The top plot shows the parameter appearing on the left of the Boussinesq criterion Eq.\eqref{eq:summaryJexplic}. The Boussinesq condition is better fulfilled for the values of the parameter smaller than one (the Chavanne \textit{et al.} data near $Ra\sim 10^{12}$ sitting below $10^{-4}$ have been obtained in liquid helium). The bottom plot shows the compensated Nusselt $Nu Ra^{-1/3}$ versus $Ra$. We note that, for $Ra \sim 10^{13}$ (vertical line), the datasets already sits on different $Nu(Ra)$ trends while some of their non Boussinesq parameters, displayed on the top plot, roughly overlap. \label{HeCritere2}}
\end{figure}

Data from \cite{Chavanne1997,Niemela2000,Niemela2003} have been corrected from the side wall effect \cite{RocheEPJB2001,AhlersWall2001,Verzicco2002} and filtered to reject those with $\alpha \Delta > 0.2$, that is a $10\%$ maximum relative variation of the density in each boundary layer (first criterion). All cells are cylindrical and their height and aspect ratio $\Gamma=diameter / height$ are reported in the legend of figure\,\ref{HeCritere2}. 

The top plot shows the LHS of the criterion Eq.\eqref{eq:summaryJexplic} versus $Ra$.  We recall here that the fit of $\theta_{rms}^{\star}$ used to derive Eq.\eqref{eq:summaryJexplic} from Eq.\eqref{eq:summaryJnew} is smooth over nearly 6 decades, up to nearly $Ra=10^{15}$ and that we assume that its extrapolation over one extra decade (up to $Ra=10^{16}$) provides a correct order of magnitude. From the plot, we find that this non-Boussinesq parameter reaches values larger than 0.2 for the highest $Ra$ ($Ra\sim 10^{16}$), which may no longuer be considered as significantly smaller than unity. 
This suggests that Boussinesq approximation could be violated for such ultra high $Ra$. 

Besides, this shows that increasing the He cell height $h$ to reach higher $Ra$ -in Boussinesq conditions- is at best efficient like $h^2$ and not $h^3$. Indeed, if we neglect the weak $Ra$ dependence of Eq.\eqref{eq:summaryJexplic} and increase by a factor -say- 10 the cell height, $\alpha$ should be 10 times smaller for  $\alpha \Delta_{h}$ to stay constant (fulfilment of Eq.\eqref{eq:summaryJexplic}). On the other hand, $\Delta$ can be made 10 times larger (fulfilment of Eq.\eqref{eq:summaryB}). In temperature-density area where these high $Ra$ data are obtained, a 10 fold decrease of $\alpha$ is roughly correlated with a 10 fold increase of the thermal diffusivity but with little change in the kinematic viscosity. All parameters considered, the maximum ``Boussinesq'' $Ra$ will only be 100 times larger.
One the other hand, this implies that such tall cells could be interesting models for atmospheric convection. Indeed, according to the estimate presented above we expect a similar violation of the Boussinesq approximation in the atmosphere. 

The bottom plot of this figure illustrates that the compensated Nusselt $Nu Ra^{-1/3}$ has different trends for $Ra>10^{12}$ as mentioned earlier in the text.
Regarding this issue, the top plot also shows that the non-Boussinesq parameter is below $10^{-2}$ for $Ra \sim 10^{13}$ and -more important- that it is approximately the same for datasets with different $Nu(Ra)$ trends (the vertical line across the plots illustrates this point). This shows that non-Boussinesq deviations of the sort encompassed by the present analysis cannot explain the discrepancy of heat transfer efficiency among He cells for $10^{12}<Ra < 10^{14}$  \cite{Wu1992, Chavanne1997,Niemela2000, RochePRE2001,Niemela2003,RochePoF2005,Niemela:2006}.

\subsection{Conclusion and Perspectives}

We derived a set of 4 criteria (Eq.\eqref{eq:summary}) for assessing the applicability of Boussinesq approximation in a fluid with constant properties at high Rayleigh number $Ra$ and intermediate $Pr$. The fourth criterion, scaling quadratically with stratification, was found to be always less stringent than the 3 others in our numerical estimations.\\

The first criterion $\alpha \Delta \ll 1$ simply reflects the incompressibility condition in the boundary layers. The 3 others are indirectly associated with stratification effects. They are written as functions of the adiabatic gradient but the thermodynamics relation Eq.\eqref{relationThermo} allows to translate them into equivalent density stratification criteria. Contrary to a widespread idea, tall cells are not necessarily more Boussinesq than small ones for given $Ra$ and $Pr$, because of these stratification effects.\\

Using fits derived from the RB literature, we found that the first criterion $\alpha \Delta \ll 1$ is the most relevant one for assessing the ``Boussinesqness'' of typical laboratory experiments conducted with water and air. In geophysical flows for which stratification occurs on very large scales, the second and third criteria could become more relevant than the first one.\\

In cryogenic helium, in addition to the first criterion, the validity of second one (Eq.\eqref{eq:summaryJnew}) deserves to be carefully  examined at ultra high $Ra$, above $10^{16}$ typically. This point and its consequences on very high $Ra$ experiments is discussed in the text. At such high $Ra$, this second criterion is a sufficient condition for density fluctuations in the bulk to be more correlated to temperature fluctuations than to pressure fluctuations (Eq.\eqref{eq:stratNew}) and to neglect, in the heat equation, a thermal heat source (or sink) term due to pressure-fluctuation-induced compression (or expansion) (Eq.\eqref{eq:condition.thermique.2.BulkNew}). 

At lower Rayleigh numbers (typically $10^{12}<Ra<10^{14} $), we find that the apparent incompatibility of heat transfer measurements reported in literature cannot be explained 
by such non-Boussinesq effects.\\

In the Boussinesq approximation, the equations of the fluid are only  parametrised by $Ra$ and $Pr$. The present work shows that a weaker approximation of the equations -preserving some terms proportional to the parameters $\alpha \Delta$ and/or $\alpha \Delta_{h}$ (or equivalently $\chi \rho g h$)- would enable to account for first deviations from the Boussinesq limit of turbulent Rayleigh-B\'enard cells filled with a fluid with constant properties. As a perspective, a DNS simulation conducted with such generalised equations at moderate turbulent Rayleigh numbers (say $10^{8}<Ra<10^{10}$) but with large enough values of the non-Boussinesq parameters $\alpha \Delta$ or $\alpha \Delta_{h}$ would provide valuable quantitative information on deviations from the Boussinesq limit
.  In particular, simulation with $\alpha \Delta_{h}$ of order unity can still be performed assuming incompressibility. Such simulations would give a precise quantitative meaning on the $\ll$ symbol used in the proposed Boussinesq criteria. 




\begin{acknowledgments}
We thank R. Verzicco for sharing DNS data, which were used to validate a scaling law for pressure fluctuations, Q. Zhou for providing some physical properties of water, J. Niemela for sending data from his aspect ratio 0.5 experiment and my colleagues B. Chabaud and H. H\'ebral for proof-reading. We acknowledge discussion with M. Pons, P. Carles and more especially with B. Castaing. This work was initiated thanks to the R\'egion Rh\^one-Alpes contract 301491302.
\end{acknowledgments}



\end{document}